\documentclass[final,1p,times]{elsarticle} % do not change
\usepackage{graphicx} % do not change
\usepackage{amssymb} % do not change
\usepackage{amsthm} % do not change
\usepackage{lineno} % do not change

\journal{Nuclear Physics A} % do not change
\begin{document} % do not change

\begin{frontmatter} % do not change

%% QM09Author: please enter your  
%% Title, author and address info here; please do not use footnotes

% Your Title - please modify
\title{Highlights from PHENIX-I: Initial State and Early Times}

% Principle author, and co-authors - please modify
\author{Michael Leitch$^{a}$ (for the PHENIX collaboration) }

% Address - please modify
% note that if you have authors from several institutions, we recommend
% labelling these [a], [b], [c] etc.
\address[a]{Los Alamos National Laboratory, % label [a]
P-25 MS H-846,
Los Alamos, NM, 87545, USA}

\begin{abstract} % do not change
%% Text of abstract goes here - please modify
We will review the latest physics developments from PHENIX concentrating
on cold nuclear matter effects, the initial state for heavy-ion collisions,
and probes of the earliest stages of the hot-dense medium created in those
collisions. Recent physics results from $p+p$ and $d+Au$ collisions; and from direct
photons, quarkonia and low-mass vector mesons in A+A collisions will be highlighted.
Insights from these measurements into the characteristics of the initial state
and about the earliest times in heavy-ion collisions will be discussed.
\end{abstract} % do not change

\end{frontmatter} % do not change

%% QM09: we keep linenumbers at least for initial version
%% \linenumbers % do not change

%% start of main text - please modify. Below is a sub-set (single section) 
%% of an earlier proceedings that show how one can handle references 
%% and figures etc.
%%\section{}\label{}

\section{Introduction}

In this overview, we will discuss selected highlights from PHENIX in the areas relating
to the initial state and early times,
focusing only on those which we believe to be the most significant new results.
These will include 1) the suppression of rapidity-separated hadron pairs in $d+Au$ collisions,
2) the contributions of quarkonia and Drell-Yan to the non-photonic
single electrons used to detect heavy quarks, 3) the continued suppression of
the $J/\psi$ in $Cu+Cu$ collisions to high-$p_T$, 4) the suppression of $\Upsilon$s in $Au+Au$ collisions,
and 5) estimates of the initial temperature from direct photons in $Au+Au$ collisions.

\section{Cold Nuclear Matter (CNM) and Gluon Saturation}

The physics that modifies hard processes in nuclei relative to those on a free nucleon,
often called cold nuclear matter (CNM) effects, includes 1) traditional shadowing either
from global fits or from coherence models, 2) gluon saturation at small momentum fraction
($x$) which is amplified in the nuclear environment, and 3) initial-state energy loss
and multiple scattering. For hadron pairs with a rapidity separation between the two
hadrons in the pair, where one "triggers" on a mid-rapidity ($|\eta| < 0.35$) hadron and studies correlations
with a forward-rapidity ($3.1 < \eta < 3.9$) hadron, there are two pictures which attempt to describe
the characteristics of the process. QCD based pictures, such as those used by Vitev~\cite{Vitev}
that include non-leading-twist shadowing, give suppression of the pairs compared to the mid-rapidity
trigger particle. An alternative approach which represents gluon saturation in the
color-glass-condensate (CGC) model~\cite{CGC} also gives suppression and gives broadening
of the angular correlation peak between the two particles in the pair. In the CGC picture, a
mono-jet mechanism becomes important, where a single jet has its momentum balanced by multiple
gluons coupling to the saturated gluon field.

%\begin{figure}[ht]
%\centering
%%
%% here is an example syntax if you want to have a single eps file in the figure
%\includegraphics[scale=0.19]{integrated_charm_with_FONLL.eps}	       
%% - then you need to use commands a la those shown in the included 'mkpaper'
%% script: e.g. by doing 'source mkpaper'
%%
%% next is an example for including 2 non-eps files in the same figure
%\includegraphics[width=0.4\textwidth]{integrated_charm_with_FONLL.jpg}
%\includegraphics[width=0.45\textwidth]{all_spectras_together.jpg}
%% in the example here we give slightly more horiz. space for the 2nd figure 
%% - then you can just use 'pdflatex example.tex', and get a pdf directly
%%
%\includegraphics[width=0.4\textwidth]{figures/pp_jpsi_sub_logo.eps}
%\includegraphics[width=0.45\textwidth]{figures/figure2_ppg94.eps}
%\caption[]{Fraction of beauty vs single-electron transverse momentum.}
%\label{beauty_fraction}
%\end{figure}

\begin{figure}[ht]
\centering
\includegraphics[width=0.49\textwidth]{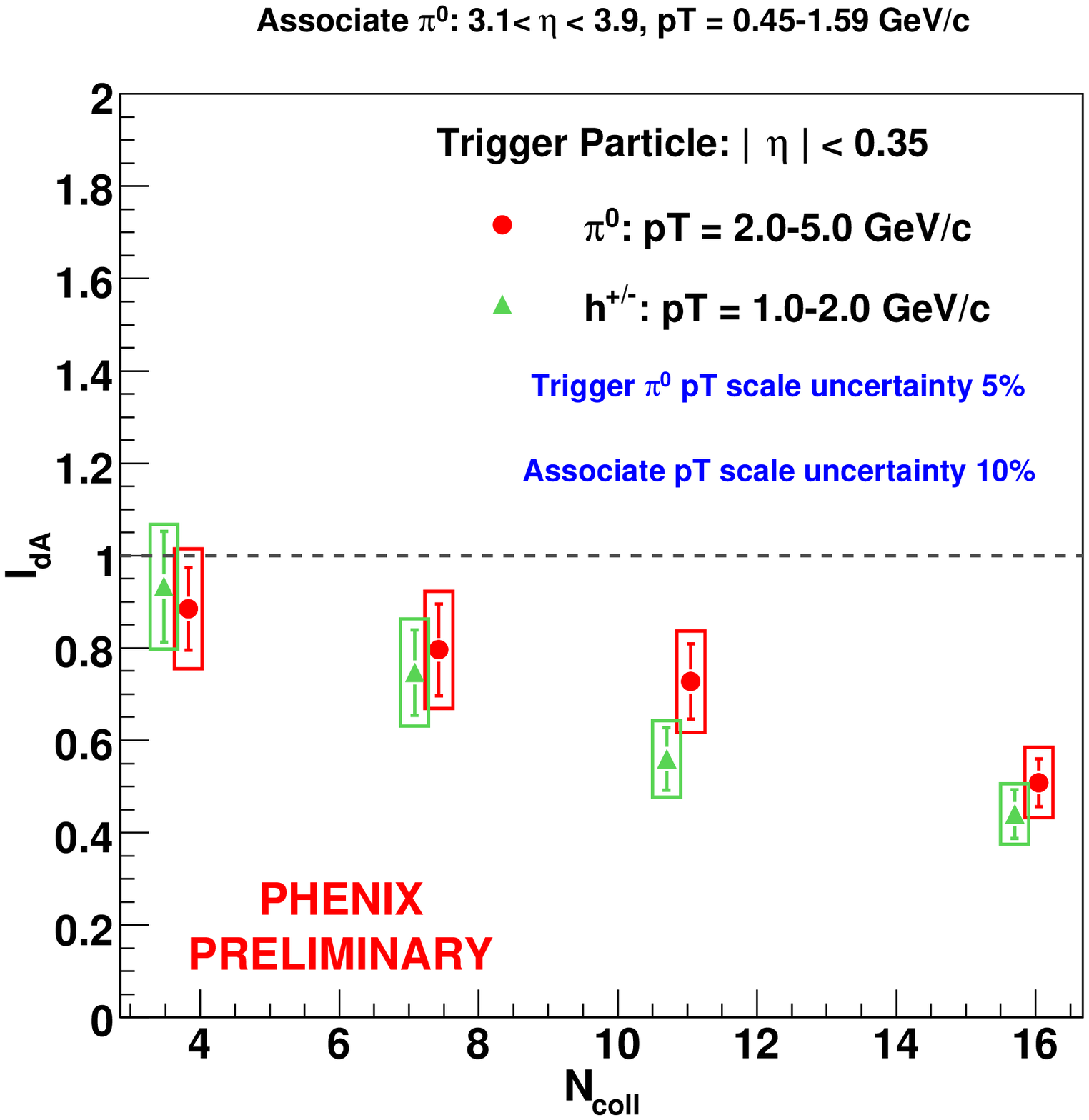}
\includegraphics[width=0.49\textwidth]{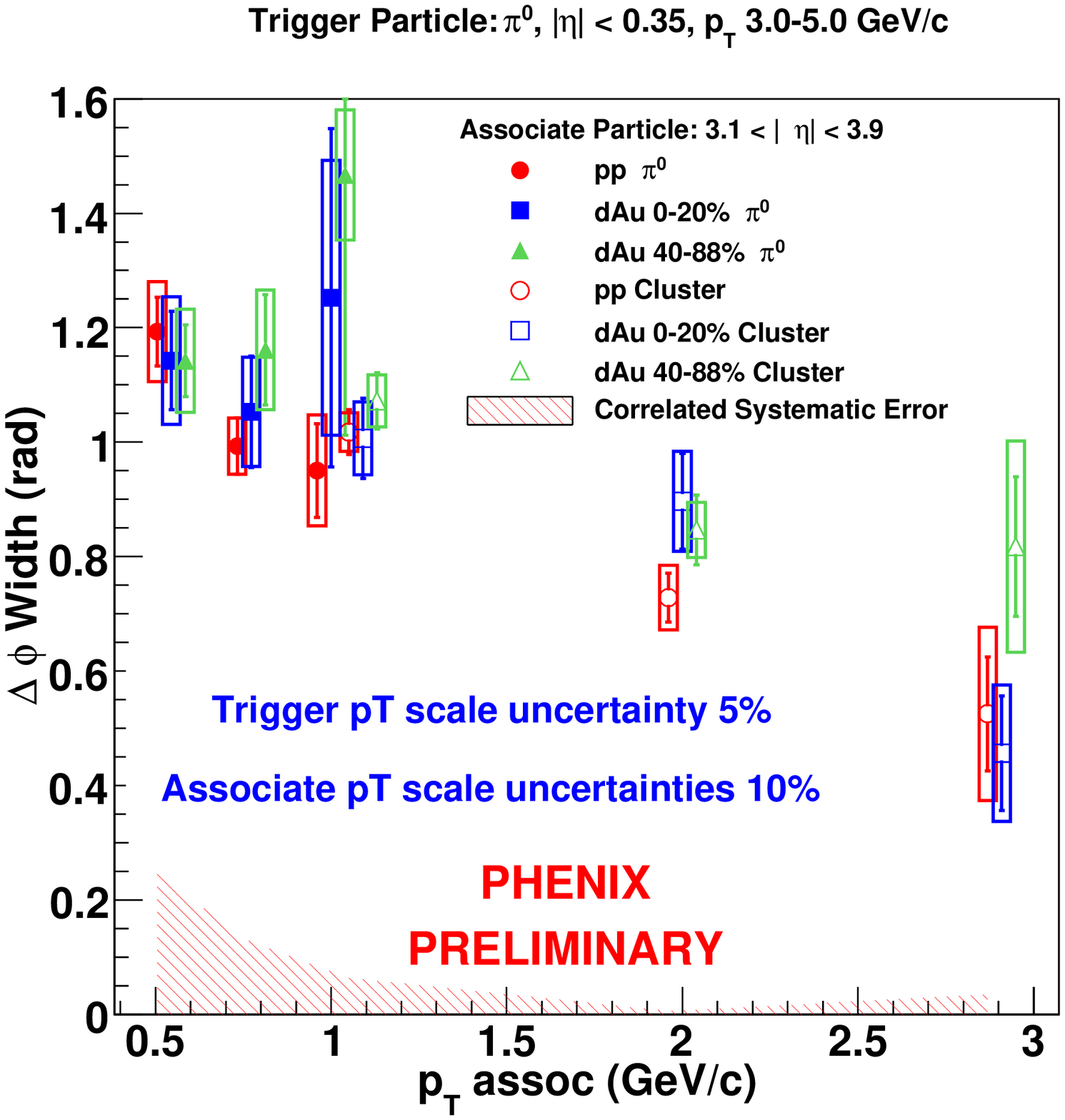}
\caption[]{(Color online) Centrality dependence of $I_{dAu}$ for rapidity-separated hadron pairs (left).
Correlation width, $\Delta\phi$, vs $p_T$ of the associated mid-rapidity ($|\eta| < 0.35$) $\pi^0$ (filled
points) for $p+p$ and for different centrality $d+Au$ collisions, showing no broadening within
the substantial uncertainties of the data points (right). Also shown are similar results for higher energy
clusters (open symbols) where $\pi^0$'s and photons are not resolved.}
\label{mpc_fig}
\end{figure}

Using the new Muon Piston Calorimeters (MPC) in PHENIX we are able to study correlations of
rapidity-separated hadron (h$^{\pm}$ or $\pi^0$) pairs, where one triggers on a hadron at
mid-rapidity and studies correlations with hadrons at forward rapidity in the MPC. For these
studies we use the ratio $I_{dAu}$, which is the pair efficiency relative to the mid-rapidity "trigger"
hadron for $d+Au$ divided by that for $p+p$,

\begin{equation}
I_{dAu} = { {N^{pair}_{d+Au}[(\eta = 3.5) + (\eta = 0)]/N^{trig}_{d+Au}(\eta = 0)}
  \over
  {N^{pair}_{p+p}[(\eta = 3.5) + (\eta = 0)]/N^{trig}_{p+p}(\eta = 0)} }
\end{equation}

\noindent
Preliminary results~\cite{beau} for the centrality dependence (in terms of number of collisions, $N_{coll}$)
of $I_{dAu}$ in Fig.~\ref{mpc_fig} (left) show increasing suppression for more central collisions.
The angular correlations of the pairs were also studied, but showed no broadening in the relative angle
$\Delta\Phi$ outside the substantial uncertainties in the present preliminary result, Fig.~\ref{mpc_fig} (right).

We have also studied hadron pairs in $d+Au$ collisions where both hadrons are at mid-rapidity, this time in terms of $J_{dAu}$
which is basically the same as $R_{dAu}$ for a single particle, but in this case for pairs,

\begin{equation}
J_{dAu} = { {(PairYield)_{dAu}} \over {<N_{coll}> (PairYield)_{pp}} }
\end{equation}

\noindent
These pairs exhibit a very large Cronin-like enhancement, i.e. they scale faster than $N_{coll}$
($J_{dAu} > 1$) and both $J_{dAu}$ and the angular correlation width decrease for larger $p_T$~\cite{jia}.

\section{Open Heavy Quarks}

\begin{figure}[ht]
\centering
\includegraphics[width=0.47\textwidth]{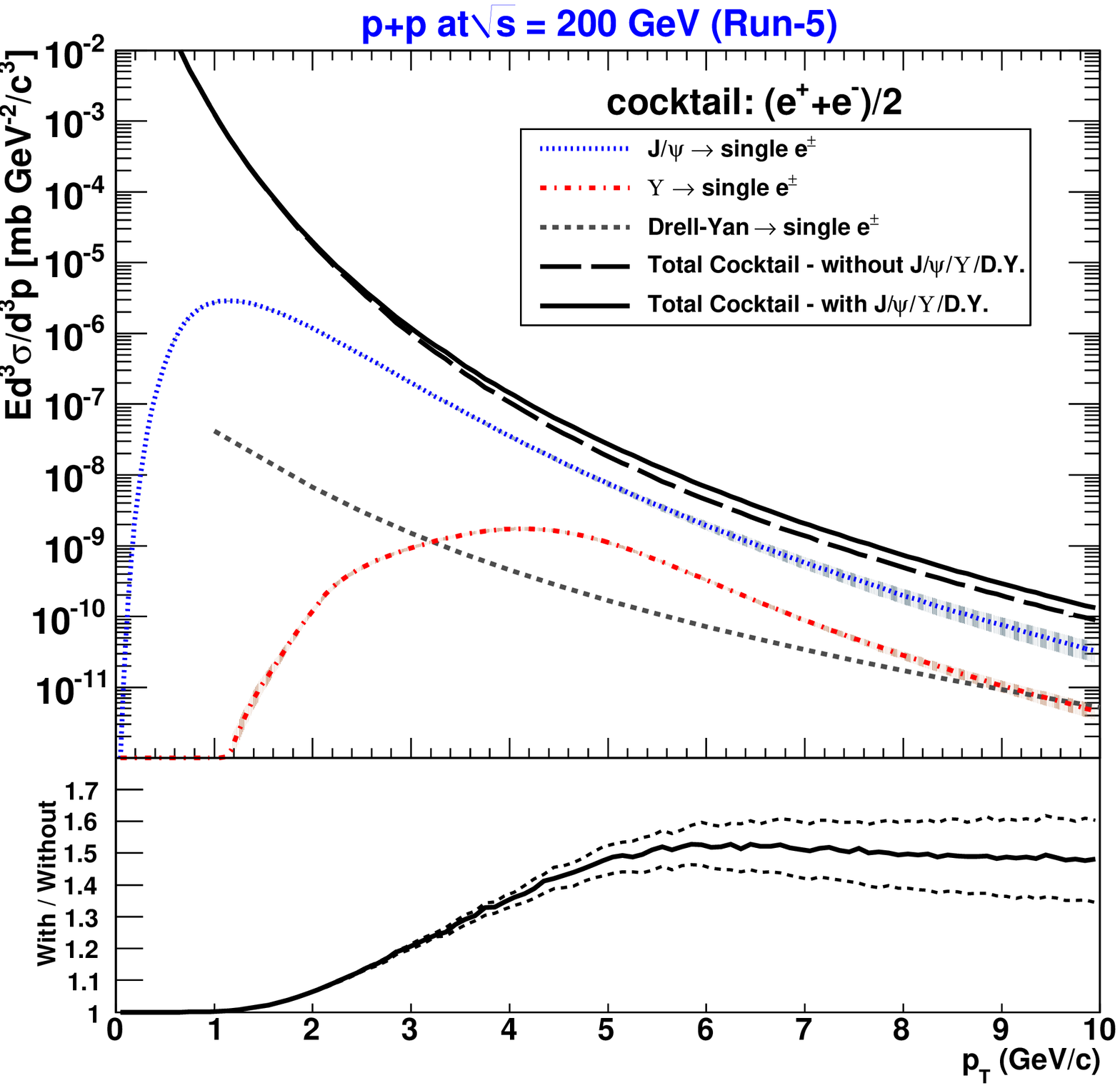}
\includegraphics[width=0.47\textwidth]{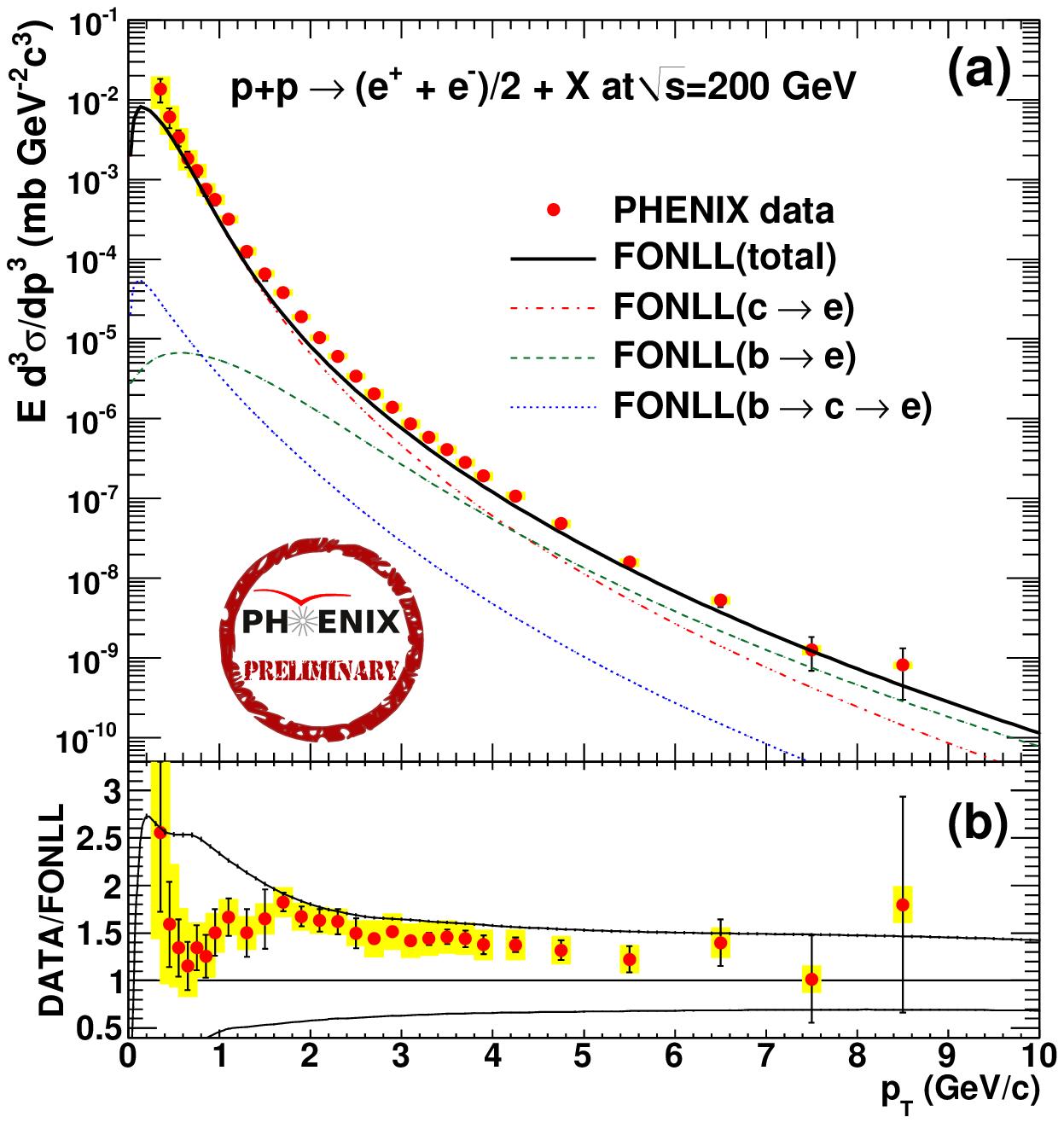}
\caption[]{(Color online) Contributions of quarkonia and Drell-Yan to the background for single electrons from heavy meson decays
and the change relative to the background before their inclusion (left).
Single electron spectrum from heavy quarks corrected for these contributions 
compared to a FONLL calculation~\cite{FONLL} (right).}
\label{quarkonia_electrons}
\end{figure}

Recent studies of the contribution of quarkonia and Drell-Yan to the spectrum of single electrons
from heavy quarks have determined that for transverse momenta above about 5\ GeV/$c$ these contributions
can amount to up to 16\% of the total non-photonic electron yield~\cite{dion}. The contributions of $J/\psi$,
$\Upsilon$, and Drell-Yan are shown in Fig.~\ref{quarkonia_electrons} (left) and one can see that the
$J/\psi$ gives the dominant contribution. After subtracting off these estimates of the
contributions to the $p+p$ collision data, with careful attention to their uncertainty,
shown in Fig.~\ref{quarkonia_electrons} (right),
the net yield of electrons from heavy quark decay has moved from a little above, to slightly below, the
FONLL model's upper uncertainty limit. Similar corrections, but with larger uncertainties have been applied
for $Au+Au$ collisions. However, because both $p+p$ and $Au+Au$ are lowered by about the same amount, the resulting
nuclear depencence in $Au+Au$ collisions, $R_{AA}$, is not significantly changed.

\begin{figure}[ht]
\centering
\includegraphics[width=0.47\textwidth]{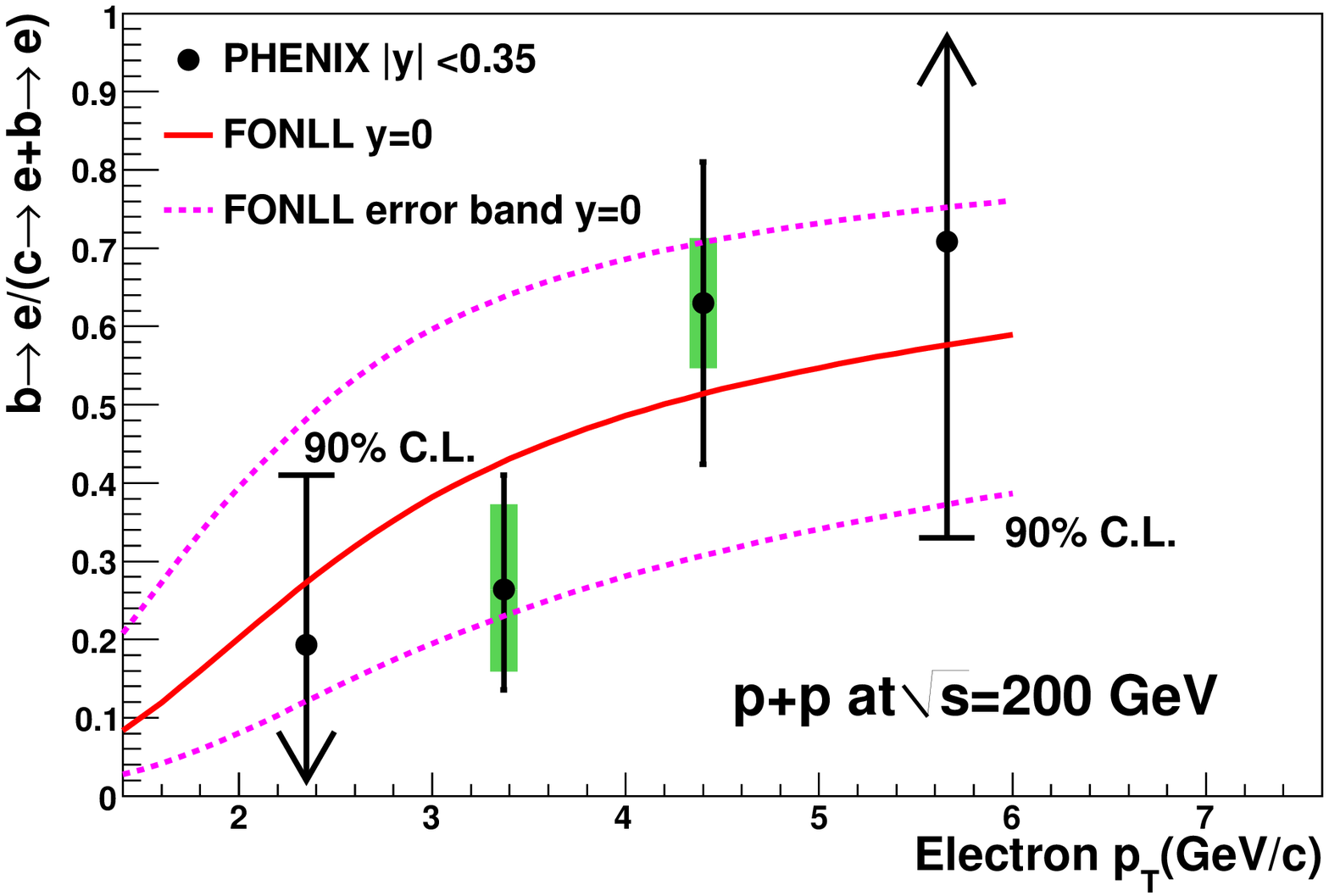}
\includegraphics[width=0.47\textwidth]{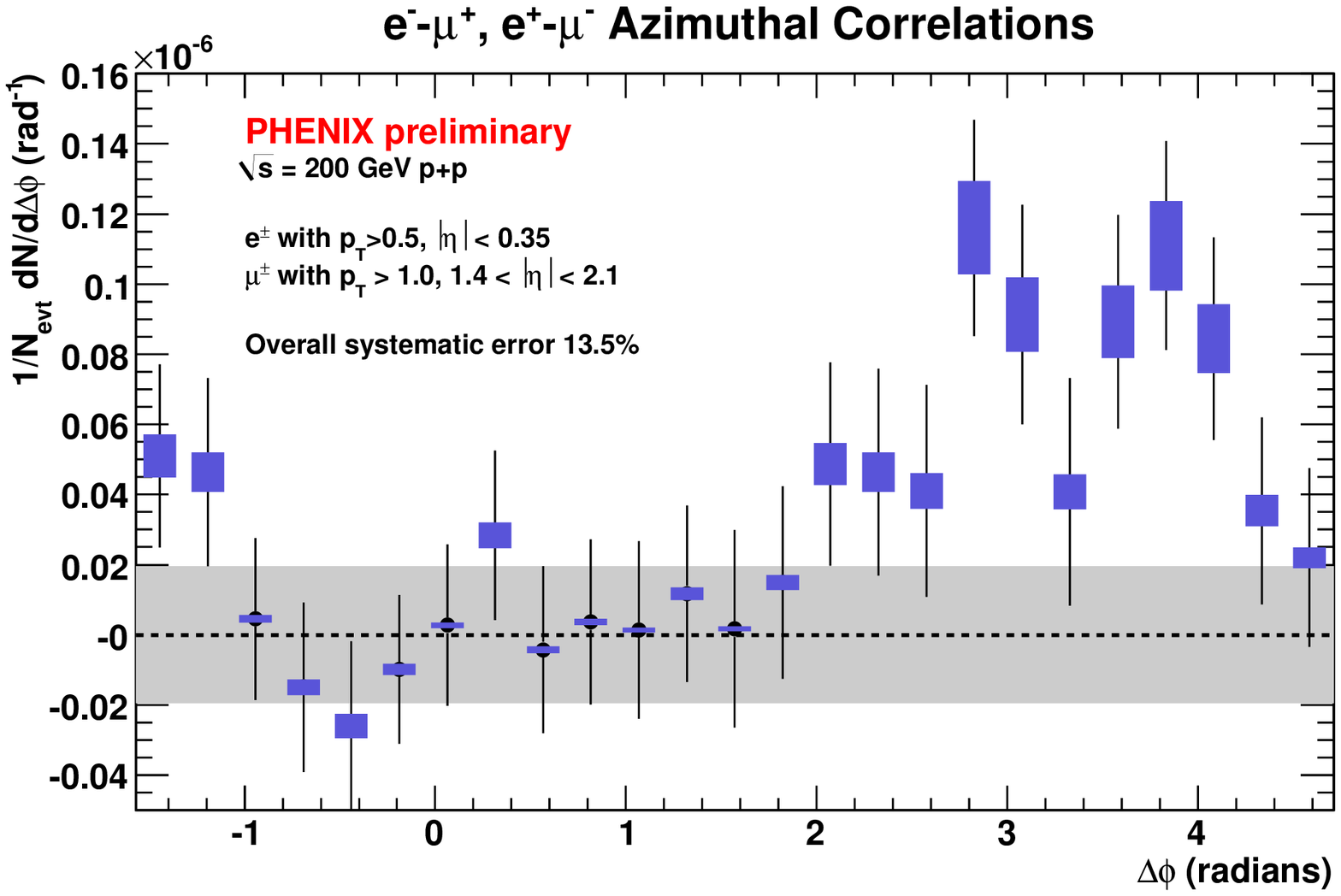}
\caption[]{(Color online) Fraction of beauty, $(b \rightarrow e)/(b+c \rightarrow e)$, vs single-electron
transverse momentum compared to FONLL calculations~\cite{FONLL} (left).
Early electron-muon pair charm signal for $p+p$ collisions (right).}
\label{beauty_fraction}
\end{figure}

Most open-heavy flavor meson measurements at RHIC to date are not able to separate
contributions to the single electrons
from charm and beauty, while theoretical predictions of energy loss and flow in the hot-dense medium
created in high-energy heavy ion collisions are generally quite different for charm and beauty. Recently
a new method has been employed, where one studies the correlations of hadrons near the observed electron
and exploits the fact that the decay of a beauty meson into an electron and hadron produces a broader
correlation and lower efficiency for observing the pair than that of a charm meson. Using this technique,
the fraction of $(b \rightarrow e)/(b+c \rightarrow e)$ has been determined~\cite{ppg094} and is shown
vs $p_T$ in Fig.~\ref{beauty_fraction} (left).

PHENIX has also measured open-heavy flavor mesons at forward rapidity via their decay to single muons, but so
far not with enough precision to define the shape of the cross section vs rapidity. However, three
different methods in PHENIX now yield consistent cross sections in $p+p$ collisions at mid rapidity: single electrons
via cocktail subtraction, a converter method, and with di-electrons. Using the beauty fraction
determined above, a beauty cross section of
$\sigma_{b{\bar b}}= 3.2 ^{+1.2}_{-1.1}(stat) ^{+1.4}_{-1.3}(sys)\ {\mu}b$
has also been determined~\cite{ppg094}.

Finally, the first proof-of-princible measurement of charm pairs via electron-muon correlations in $p+p$ collisions
has been made~\cite{tatia}
and is shown in Fig.~\ref{beauty_fraction} (right). The peak at $\pi$ radians in $\Delta\Phi$ is from these
correlated pairs. This method promises to provide another independent measurement of charm in the near future,
as luminosities increase and allow substantial yields for this rare signal.

\section{Quarkonia Production and Suppression}

\begin{figure}[ht]
\centering
\includegraphics[width=0.47\textwidth]{figures/lambda_pt2_theory.eps}
\includegraphics[width=0.49\textwidth]{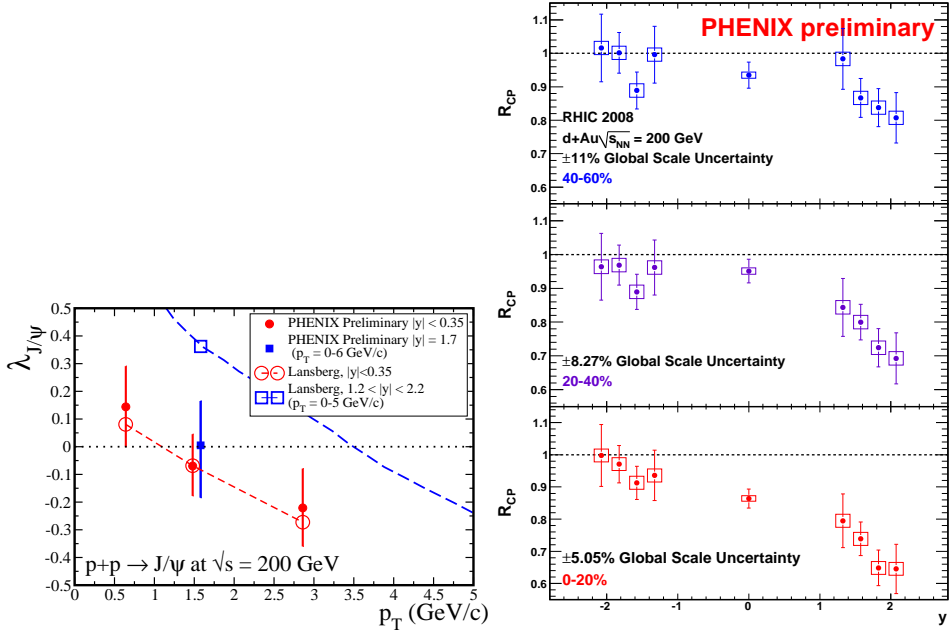}
\caption[]{(Color online) Polarization vs $p_T$ in the helicity frame for $J/\psi$ production in 200\ GeV $p+p$ collisions
with mid-rapidity points as red circles and forward rapidity points as blue squares (left).
$R_{CP}$ vs rapidity for $J/\psi$ production in 200\ GeV $d+Au$ collisions for three different
centrality bins, with the most central collisions (0-20\%) on the bottom (right).}
\label{jpsi_polarization}
\end{figure}

The simultaneous theoretical description of both the cross section and the polarization of the $J/\psi$ in
hadron production has long been a challenge. A new analysis of the 2006 PHENIX $p+p$ data agrees well with the
previous results, has significantly higher precision, and agrees well with the Lansberg s-channel cut color-singlet
model~\cite{lansberg}. The decay polarization of the $J/\psi$ measured by PHENIX at mid
and forward rapidity is shown in Fig.~\ref{jpsi_polarization} (left), where the Lansberg model reproduces the small
polarization falling with $p_T$ at mid rapidity (red points), but predicts a larger polization than the null
polarization seen at forward rapidity (by 2-3 sigma)~\cite{daSilva}. Improved polarization measurements at
forward rapidity in several bins in $p_T$ are expected soon, and may help clarify the situation.

New results for the $J/\psi$ from the 2008 $d+Au$ run with approximately thirty times larger integrated luminosity
than that of the previous (2003) $d+Au$ results are beginning to emerge, with the first preliminary result in terms
of $R_{CP}$,

\begin{equation}
R_{CP}^{0-20\%} = { { N_{inv}^{0-20\%}/<N_{coll}^{0-20\%}> } \over { N_{inv}^{60-88\%}/<N_{coll}^{60-88\%}> } },
\end{equation}

\noindent
shown in Fig.~\ref{jpsi_polarization} (right) vs rapidity for three different centrality
bins~\cite{daSilva,darren}. One sees essentially no
nuclear dependence at backward rapidity, a little at mid rapidity, and increasing suppression with
centrality at forward rapidity in the nuclear shadowing region (large rapidity corresponds to small
momentum fraction down to about $x = 2 \times 10^{-3}$ and is in the shadowing region).
PHENIX is working on more comprehensive results
for the near future in terms of $R_{dAu}$, the nuclear dependence relative to $p+p$ - the much higher statistical
precision of this new data requires precision systematics and more careful analysis.

\begin{figure}[ht]
\centering
\includegraphics[width=0.47\textwidth]{figures/rcucu_pt.eps}
\includegraphics[width=0.47\textwidth]{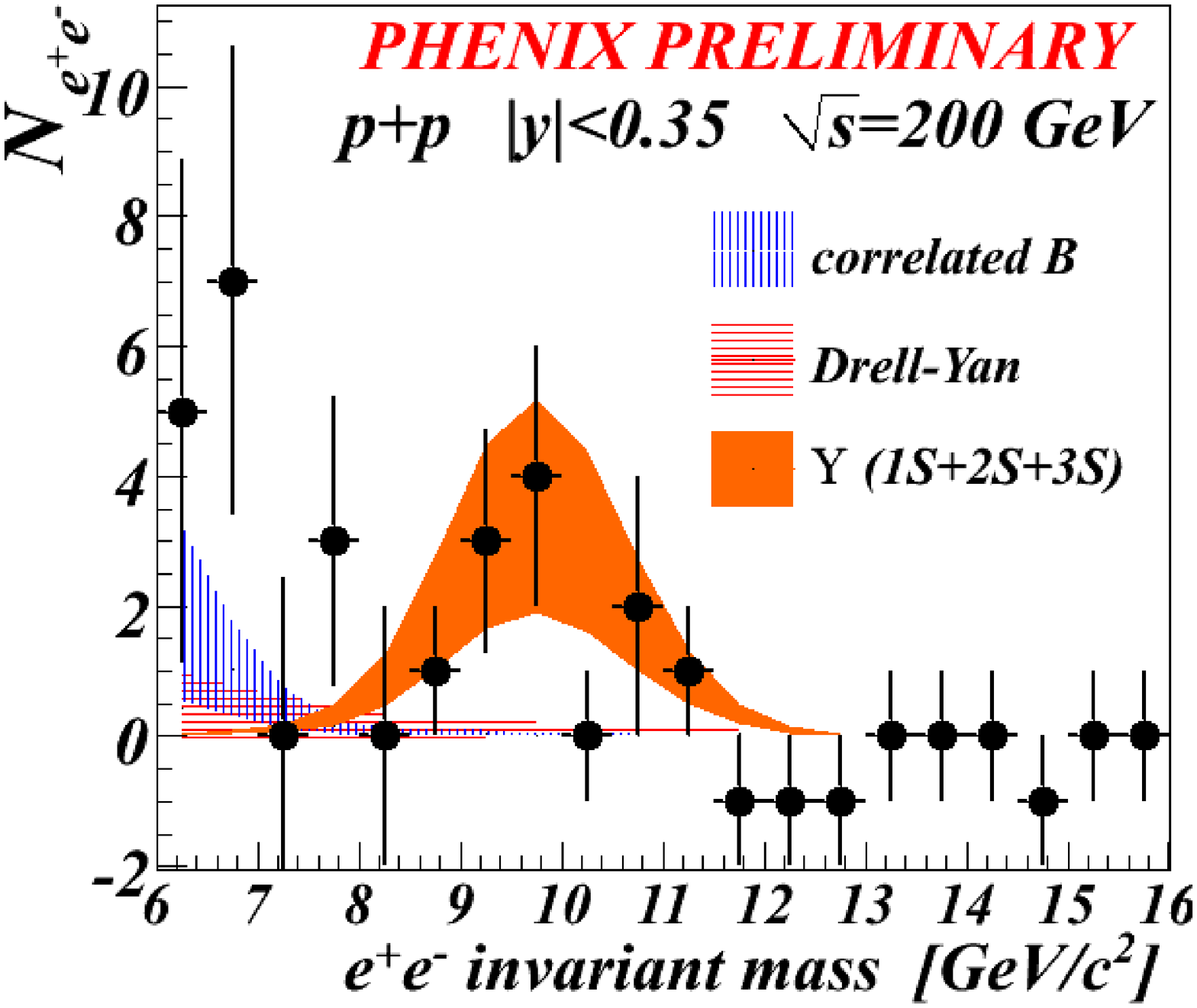}
\caption[]{(Color online) $R_{CuCu}$ vs $p_T$ for mid-rapidity $J/\psi$s out to 9\ GeV/$c$ compared to several theoretical
models~\cite{hot_wind,highpt_theory} (left).
Invariant mass spectrum at mid rapidity for $p+p$ collisions at high mass showing the $\Upsilon$ family and
other components of the spectrum.}
\label{ups_figs}
\end{figure}

New preliminary results for $J/\psi$ $R_{AA}$ in $Cu+Cu$ collisions show continuing suppression up to at least
7\ GeV/$c$ in $p_T$. In Fig.~\ref{ups_figs} (left) this suppression is compared to several theoretical models, including
the "hot-wind" AdS/CFT inspired model~\cite{hot_wind} which is inconsistent with the data. Eventually, due to the
Cronin effect seen in $d+Au$ collisons, which causes a change from suppression to
enhancement at high-$p_T$, one would expect $R_{AA}$ to return to unity at large $p_T$, but there is no evidence
of that yet from these results.

\begin{figure}[ht]
\centering
\includegraphics[width=0.47\textwidth]{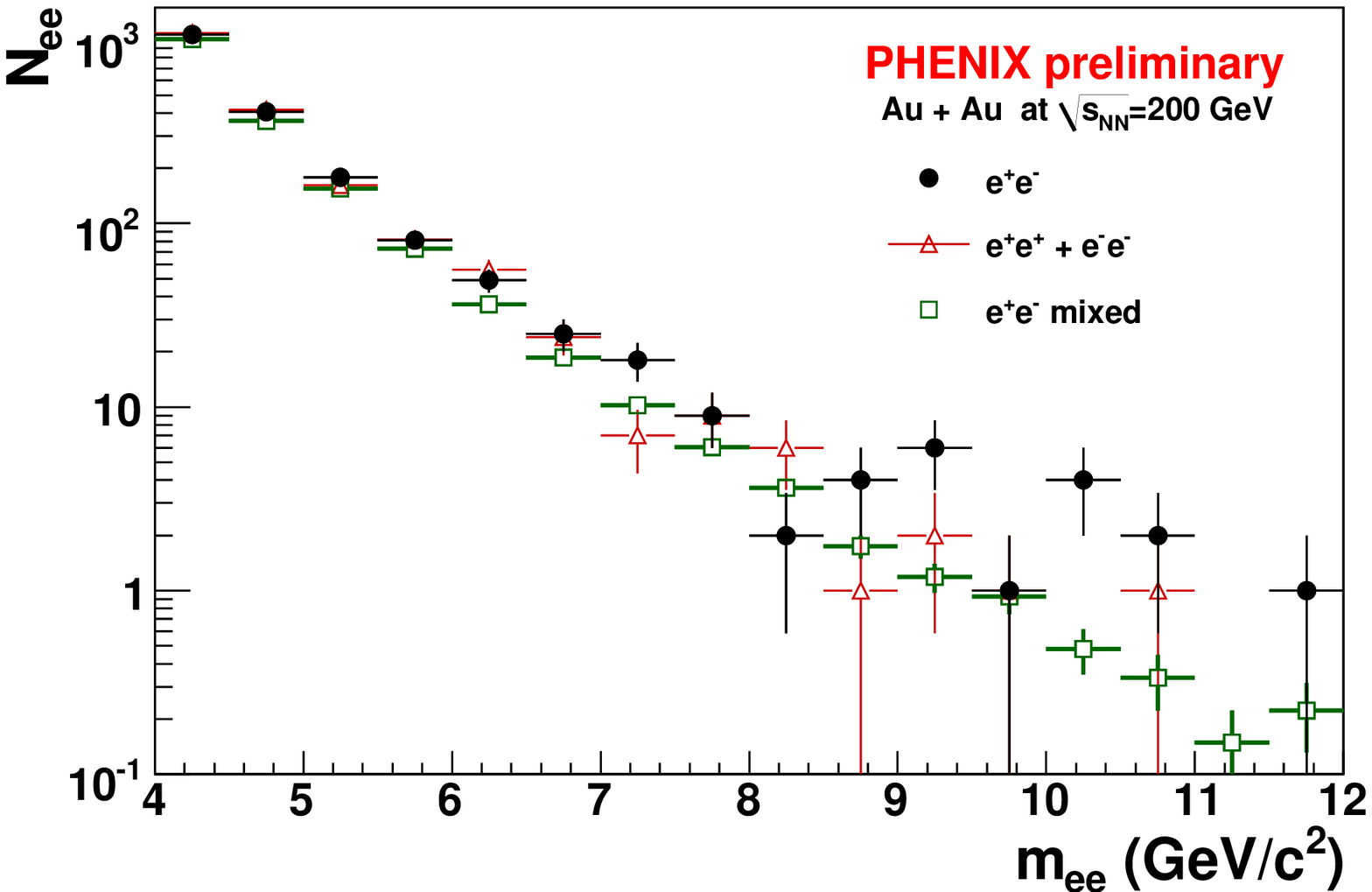}
\includegraphics[width=0.47\textwidth]{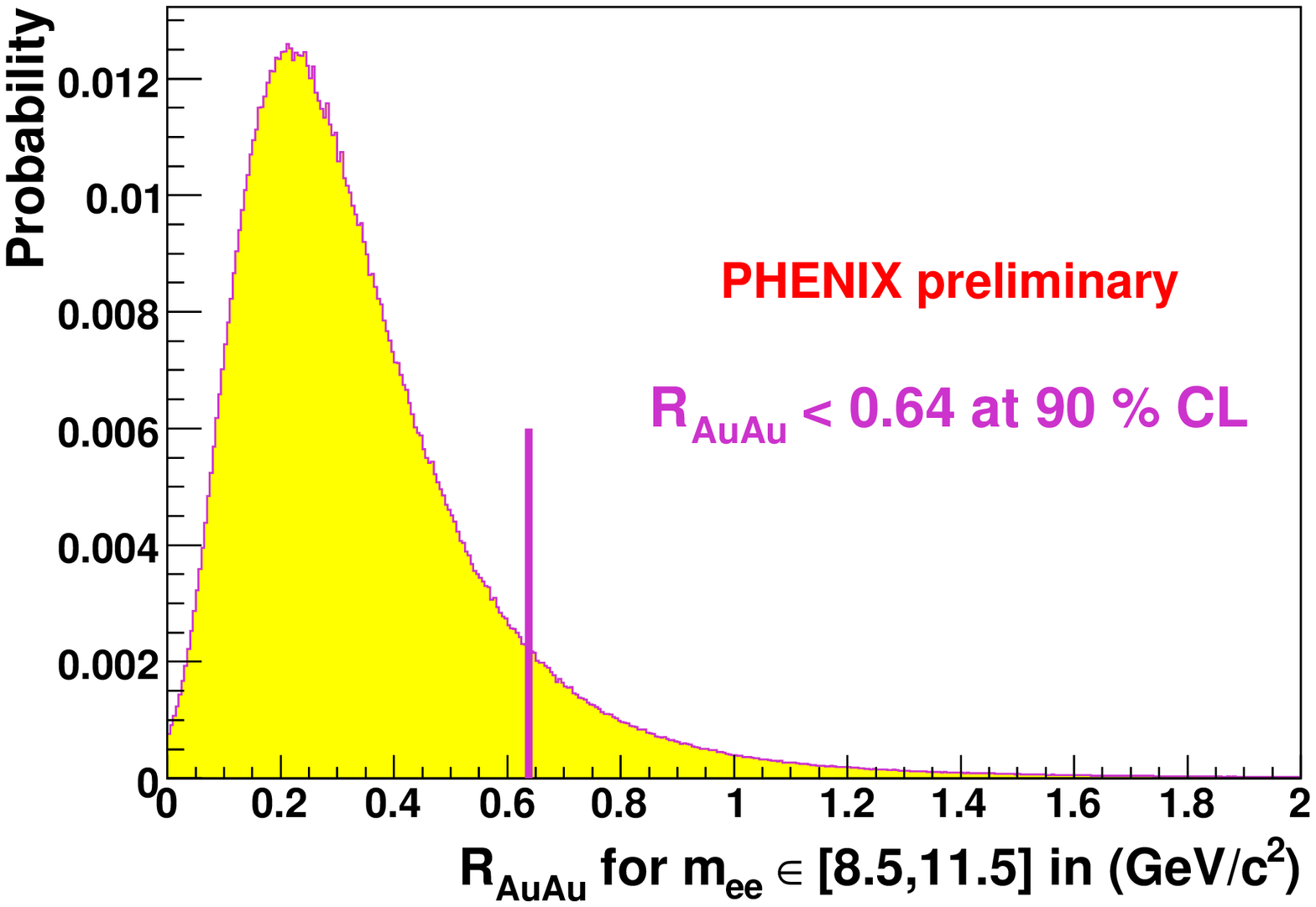}
\caption[]{(Color online) Di-electron pair mass at mid rapidity for 200 GeV Au+Au collisions in the $\Upsilon$ mass region, where $e^+e^-$
pairs are shown as black points, like-sign pairs as red points, and mixed $e^+e^-$ background pairs in green (left).
The probability distribution vs $R_{AuAu}$ determined from the $Au+Au$ and $p+p$ data (right).}
\label{ups_figs2}
\end{figure}

With the increasing luminosities provided by the RHIC machine, PHENIX is now beginning to accumulate useful
number of $\Upsilon$s for various kinds of collisions. From the 2006 $p+p$ run, as shown in
Fig.~\ref{ups_figs} (right), we now have a preliminary
cross section for dielectron events in the $\Upsilon(1S+2S+3S)$ mass region [8.5,11.5 GeV/$c^2$] of
$BR*d\sigma/dy\ (|y|<0.35) = 114^{+46}_{-45}\ pb$ . A small number of dielectron pairs from Drell-Yan and from
open beauty pairs may also contribute in that mass region, but this contribution is estimated to be less
than 15\% and is included in the systematic uncertaintiy.
Using a similar signal for $Au+Au$ collisions, shown in Fig.~\ref{ups_figs2} (left),
and doing a very careful statistical analysis which takes
into account the small numbers of counts in both the $Au+Au$ and $p+p$ $\Upsilon$ mass regions, we have obtained
the probability distribution for $R_{AuAu}$ in this mass region shown in Fig.~\ref{ups_figs2} (right).
From this an upper limit of $R_{AuAu} < 0.64$ at 90\% C.L. is determined~\cite{atmossa}.

Although $\Upsilon$s have long been touted as the standard candle for the melting of quarkonia in the
Quark Gluon Plasma (QGP), i.e. that they would not be screened up to very high temperatures, it is clear that
there are a number of simple non-QGP effects that could easily cause a suppression at or below the upper limit
determined above. These include 1) the suppression of $\Upsilon$ states seen in fixed target experiments~\cite{e772}
which would give about ~$0.81^2$ in $R_{AuAu}$, 2) the fact that only about 52\% of the $\Upsilon_{1S}$ do not
come from feeddown from the higher mass (2S, 3S) $\Upsilon$ states and B decays~\cite{upsilon_feeddown}, and
3) that we do not resolve the three $\Upsilon$ states (1S+2S+3S) and the 1S is only about 73\% of
the total~\cite{upsilon_fractions}.

\section{Initial State and Temperature}

\begin{figure}[ht]
\centering
\includegraphics[width=0.47\textwidth]{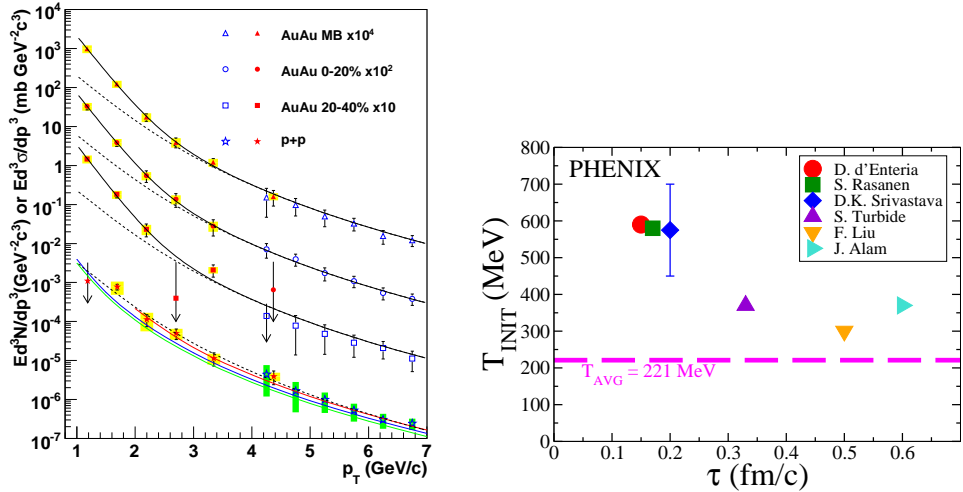}
\includegraphics[width=0.47\textwidth]{figures/tzt.eps}
\caption[]{(Color online) Direct photon spectra vs $p_T$ in $p+p$ and $Au+Au$ collisions at 200\ GeV, with the scaled
$p+p$ reference compared to the data from various centrality $Au+Au$ data as a black dashed line.
The enhancement over the $p+p$ reference at small $p_T$ for $Au+Au$ collisions is fit to an exponential (solid
black lines) to extract an inverse slope and an effective temperature (left).
Estimates of the initial temperature of the hot-dense medium from several
different models for the expansion of that medium, given the average temperature of 221\ MeV
determined from the central $Au+Au$ spectra (right).}
\label{photon_figs}
\end{figure}

Direct photon production in nucleus-nucleus collisions, although a difficult measurement, is a clean probe
of the initial-state gluon distributions in the colliding nuclei. The latest measurements in PHENIX
show no modification relative to $p+p$ collisions except for transverse momenta above about 12\ GeV/$c$. These modifications
are likely due to CNM effects (Cronin) and isopin (neutrons vs protons)~\cite{vitev_photon}.

A new method has been used recently to extract the yield of photons in the low-$p_T$ thermal region
where the production of these photons is inferred from the low-mass ($M_{ee} <\ $300\ GeV/$c^2$)
$1 < p_T <\ $5\ GeV/$c$ $e^+e^-$ spectrum~\cite{yamaguchi}.
These low-mass photons show an enhancement over the scaled $p+p$
reference, as shown in Fig.~\ref{photon_figs} (left). If interpreted as thermal photons from the hot-dense medium
and fit to an exponential slope, an average temperature of the medium (for central collisions) of
$T_{avg} = 221 \pm 23 \pm $18\ MeV is obtained. Since this is the average over the expansion, one can ask
within various theoretical descriptions for that expansion, what the initial temperature is.
Fig.~\ref{photon_figs} (right) shows various initial temperatures vs the formation time assumed in each
theoretical picture. All models indicate an initial temperature of at least 300\ MeV, well above the predicted
QGP phase transition at 170\ MeV.

\section{Summary}

We highlight here some of the recent PHENIX results that we believe to be most interesting
in areas that relate to the initial state and early times, including:
\begin{enumerate}
\item Quarkonia contribute substantially to the electrons from heavy flavor for $p_T >\ $5\ GeV/$c$ and should
be taken into account when comparing to theoretical predictions.
\item Beauty decays give 50\% or more of the single electrons for $p_T >\ $4\ GeV/$c$, so any differences between
beauty and charm for energy loss and flow may become apparent at these $p_T$ values.
\item $J/\psi$ polarization measurements at mid rapidity agree with the Lansberg color singlet model,
but at forward rapidity the $p_T$-integrated value does not.
\item For $Cu+Cu$ collisions, $J/\psi$s continue to be strongly supressed up to $p_T \simeq\ $8\ GeV/$c$.
\item Events in the ${\Upsilon}(1S+2S+3S)$ mass region at mid rapidity are suppressed in $Au+Au$ collisions by at
least 36\%, but this is not unexpected
given cold nuclear matter effects and the likely strong suppression for central $Au+Au$ collisions of the higher mass
$\Upsilon$ states.
\item Direct photons measured in the thermal region for central $Au+Au$ collisions indicate intial temperatures of
at least 300 MeV, well above the expected QGP phase transition.
\end{enumerate}

%% end of main text

%\section*{Acknowledgments} % please check/modify, comment out or delete if not needed
%This is where one places acknowledgments for funding bodies etc., if needed.
%For the large collaborations, this is listed once and for all, together with 
%the author lists etc. in the proceedings back-material.

 % do not change 
\end{document}